\documentclass[letterpaper,12pt]{JHEP3} 
\usepackage{bm}
\usepackage{amsmath}
\usepackage{yfonts}
\usepackage{epsfig,latexsym}
\newcommand{\qn}{\textswab{q}}
\newcommand{\wn}{\textswab{w}}
\newcommand{\<}{\langle}
\renewcommand{\>}{\rangle}
\renewcommand{\d}{\partial}
\newcommand{\N}{{\cal N}}

\newcommand{\q}{\bm{q}}
\newcommand{\x}{\bm{x}}
\renewcommand{\Re}{\mathrm{Re}\,}

\newcommand{\stru}{\rule[-.2in]{0in}{.2in}}
\newcommand{\struu}{\rule[-.3in]{0in}{.3in}}

\title{From AdS/CFT correspondence to hydrodynamics. II. Sound waves}

\author{Giuseppe Policastro\\
Department of Applied Mathematics and Theoretical Physics,  
Wilberforce Road, Cambridge CB3 0WA,  UK \\
Email: \email{G.Policastro@damtp.cam.ac.uk}
}
\author{Dam T.~Son and Andrei O.~Starinets\\
Institute for Nuclear Theory, University of Washington,
Seattle, WA 98195, USA\\
Emails: \email{son@phys.washington.edu, starina@phys.washington.edu}
}

\preprint{ DAMTP-2002-129 \\ INT-PUB 02-47 \\ hep-th/0210220}

\date{October 2002}

\abstract{As a non-trivial check of the non-supersymmetric
gauge/gravity duality, we use a near-extremal black brane background
to compute the retarded Green's functions of the stress-energy tensor
in $\N=4$ super-Yang-Mills (SYM) theory at finite temperature. For the
long-distance, low-frequency modes of the diagonal components of the
stress-energy tensor, hydrodynamics predicts the existence of a pole
in the correlators corresponding to propagation of sound waves in the
$\N=4$ SYM plasma.  The retarded Green's functions obtained from
gravity do indeed exhibit this pole, with the correct values for the
sound speed and the rate of attenuation.}

\begin{document}
\section{Introduction}

Any finite-temperature medium conducts sound.  This simple fact
provides a nontrivial method to check the conjecture of gauge
theory/gravity correspondence~\cite{Maldacena} at finite temperatures.
Indeed, if one can compute the correlators of the components of the
stress-energy tensor $T^{\mu\nu}$ using an AdS/CFT
prescription~\cite{GKP,Witten1}, one should recognize in them a pole
corresponding to sound-wave propagation, expected from
hydrodynamics~\cite{Landafshitz6}.

The techniques necessary for performing this check have been developed
in refs.~\cite{gamma_paper,Policastro:2002se}.  In
ref.~\cite{gamma_paper} we formulated a prescription for computing the
retarded Green's functions from gravity.  In
ref.~\cite{Policastro:2002se} we use this prescription to compute the
correlators of the shear (non-propagating) modes, and show that the
results agree with hydrodynamic expectation.  In this paper, we
conclude the program by finding the propagating sound waves from
gravity.

The paper is organized as follows.  In section~\ref{sec:field} we
discuss the Ward identities and show how leading behaviors of the
Green's functions in the infrared are completely determined.
Section~\ref{sec:gravity} is devoted to the calculation of the Green's
functions from gravity.  We show that these correlators have the same
form as expected from hydrodynamics.  Section~\ref{sec:concl} contains
concluding remarks.  The Appendix deals with the gauge invariance of
the gravitational action.

\section{Field theory (hydrodynamic) picture}
\label{sec:field}

Let us first recall (see e.g., section 2 of
ref.~\cite{Policastro:2002se}) that hydrodynamics predicts that some
elements of the stress-energy tensor have correlators with a
sound-wave pole at
\begin{equation}\label{hydro_sound}
  \omega(q) = v_s q - \frac i2 \frac1{\epsilon+P}\left(
  \zeta + \frac43\eta\right) q^2,
  \qquad v_s^2 = \frac{\d P}{\d\epsilon}\,,
\end{equation}
where $\epsilon$, $P$, $\eta$ and $\zeta$ are the energy density,
pressure, shear and bulk viscosities, respectively.  In conformal
theories $\epsilon=3P$, $v_s=1/\sqrt3$, and $\zeta=0$.  We shall now
show that the infrared form of the correlators relevant for sound
waves can be completely found from the Ward identities.  Our treatment
closely follows ref.~\cite{Yaffe}, with the addition of conformal Ward
identities.

To derive the Ward identities it is convenient to temporarily consider
the theory in curved space-time and set the space-time to flat at the
end.  For definiteness, we first consider Euclidean correlators, and
perform analytical continuation afterwards.  By considering our field
theory in a general metric, we can define a generating function
$W[g_{\mu\nu}]$ so that the correlators of $T^{\mu\nu}$ can be found
by differentiating $W$ with respect to $g_{\mu\nu}$.  
In particular\footnote{Note that the correlation functions defined
by eq. (\ref{tensor_densities}) are tensor densities, not tensor fields.} ,
\begin{equation}
  \< T^{\mu\nu}(x) \> = \frac{\delta W[g]}{\delta g_{\mu\nu}(x)}
  \bigg|_{g_{\mu\nu}=\eta_{\mu\nu}}
  \,, \qquad
  \< T^{\mu\nu}(x) T^{\lambda\rho}(y) \> = 
    \frac{\delta^2 W[g]}{\delta g_{\mu\nu}(x)\,\delta g_{\lambda\rho}(y)}
  \bigg|_{g_{\mu\nu}=\eta_{\mu\nu}}\,.
\label{tensor_densities}
\end{equation}

Assuming $W$ to be invariant under general coordinate transformations,
\begin{equation}
  W[g_{\mu\nu}] = W[g_{\mu\nu} - \nabla_\mu \xi_\nu - \nabla_\nu\xi_\mu]
    + O(\xi^2)\,,
  \label{diff_inv}
\end{equation}
by taking variation over $\xi_\mu$, we find
\begin{equation}
  \nabla_\mu \< T^{\mu\nu} \> |_{g_{\mu\nu}} \equiv
  \nabla_\mu \left\<\frac{\delta W[g]}{\delta g_{\mu\nu}}\right\>
  = 0 \,.
  \label{Ward-diff1}
\end{equation}
Putting the metric to be flat, $g_{\mu\nu}=\eta_{\mu\nu}$,
eq.~(\ref{Ward-diff1}) becomes $\d_\mu\<T^{\mu\nu}\>=0$, which is
conservation of energy and is trivially satisfied in thermal
equilibrium where $T^{\mu\nu}$ is constant.  Differentiating
eq.~(\ref{Ward-diff1}) once more with respect to $g_{\lambda\rho}$,
taking into account that $\nabla_\mu$ depends (via the Christoffel
symbols) on the metric, and then setting $g_{\mu\nu}=\eta_{\mu\nu}$, one
finds
\begin{equation}
  q_\mu\left( G_E^{\mu\nu\lambda\rho}(q)
  + \eta^{\nu\lambda}\<T^{\mu\rho}\>
  + \eta^{\nu\rho}\<T^{\mu\lambda}\>
  - \eta^{\mu\nu}\<T^{\lambda\rho}\>\right) = 0\,,
  \label{Ward_diff2}
\end{equation}
where $G_E^{\mu\nu\lambda\rho}(q)$ is the Euclidean Green's function
in momentum space,  
\begin{equation}
  G_E^{\mu\nu\lambda\rho}(q) =
  \int\!d^4x\,e^{-iq\cdot x}\< T^{\mu\nu}(x) T^{\lambda\rho}(0)\> \,.
\end{equation}
Performing analytic continuation to Minkowski space, assuming that $G_E(q)$
becomes a Minkowski Green's function $-G(q)$ (i.e., up to a sign
change), one finally obtains the following Ward
identity:
\begin{equation}
  q_\mu\left(G^{\mu\nu\lambda\rho}(q)
  - \eta^{\nu\lambda}\<T^{\mu\rho}\>
  - \eta^{\nu\rho}\<T^{\mu\lambda}\>
  + \eta^{\mu\nu}\<T^{\lambda\rho}\>\right) = 0\,.
  \label{Ward_diff}
\end{equation}

We note here the appearance of the contact terms in
eq.~(\ref{Ward_diff}).  In the supersymmetric vacuum all these contact
terms vanish, hence eq.~(\ref{Ward_diff}) becomes $q_\mu
G^{\mu\nu\lambda\rho}=0$.  However, at finite temperatures the contact
terms are present.

The conformal invariance of the $\N=4$ SYM theory leads to an
additional Ward identity.  Instead of eq.~(\ref{Ward-diff1}), one
writes
\begin{equation}
  \<{T^{\mu}}_\mu\> \equiv g_{\mu\nu} \left\< 
    \frac{\delta W[g]}{\delta g_{\mu\nu}} \right\> = {\cal O}(R^2)\,,
  \label{conf_inv1}
\end{equation}
where ${\cal O}(R^2)$ is the conformal anomaly that is quadratic in
the elements of the Riemann tensor.  Now differentiating
eq.~(\ref{conf_inv1}) with respect to $g_{\mu\nu}$, and then putting 
$g_{\mu\nu}=\eta_{\mu\nu}$, one obtains, after going to Minkowski
space,
\begin{equation}
  \eta_{\mu\nu}G^{\mu\nu\lambda\rho}(q) = 2 \<T^{\lambda\rho}\> \,.
  \label{Ward_conf}
\end{equation}
Again, the contact term in eq.~(\ref{Ward_conf}) vanishes in the
supersymmetric vacuum, but is nonzero at finite temperature.
Equations (\ref{Ward_diff}) and (\ref{Ward_conf}) are the Ward
identities we sought to establish.

We now show that the Ward identities can be used to completely
determine the forms of the correlators in the hydrodynamic limit.
Assuming, without loss of generality, that $\q$ is aligned along the
$z$ axis, $\q=(0,0,q)$, one can classify the elements of the
stress-energy tensor with respect to the ${\rm O}(2)$ rotation around
the $z$ axis.  In this paper we are interested only in the components
of $T^{\mu\nu}$ that are invariant under this rotation: $T^{tt}$,
$T^{tz}$, $T^{zz}$, and $T^{aa}=T^{xx}+T^{yy}$.  Therefore, there are
10 independent correlators: $G^{AB}$, where $A$ and $B$ can be $tt$,
$tz$, $zz$, or $aa$.  The diffeomorphism Ward identities are
\begin{equation}\label{Ward1}
\begin{split}
  - \omega (G^{tttt} + \epsilon) + q G^{tttz} &= 0\,,\qquad ~\,\,\,
  - \omega G^{tttz} + q (G^{ttzz} + \epsilon) = 0\,,\\
  - \omega G^{tttz} + q (G^{tztz} + P) &= 0\,,\qquad\,\,\,\,
  - \omega (G^{ttzz} - P) + q G^{tzzz} = 0\,,\\
  - \omega (G^{tztz} - \epsilon) + q G^{tzzz} &= 0\,,\qquad \,\,\,
  - \omega G^{tzzz} + q (G^{zzzz} - P) = 0\,,\\
  - \omega (G^{ttaa} - 2P) + q G^{tzaa} &= 0\,, \qquad
  - \omega G^{tzaa} + q (G^{zzaa} + 2P) = 0 \,.  
\end{split}
\end{equation}
The arguments of $G$'s in eqs.~(\ref{Ward1}), as well as in
(\ref{Ward3}) below, are $\omega$, $q$. The conformal Ward identities are
\begin{equation}
\begin{split}\label{Ward3}
  - G^{tttt} + G^{ttzz} + G^{ttaa} &= 2 \epsilon\,,\quad \,\,\,\,\,
  - G^{tttz} + G^{tzzz} + G^{tzaa} = 0\,,\\
  - G^{ttzz} + G^{zzzz} + G^{zzaa} &= 2P\,,\quad
  - G^{ttaa} + G^{zzaa} + G^{aaaa} = 4P \,.
\end{split}
\end{equation}

The presence of contact terms in eqs.~(\ref{Ward1}) implies that the
Green's function $G$ obtained from the generating functional $W[g]$
does not concide with the retarded Green's function, which is defined
as
\begin{equation}\label{GRdef}
  G_R^{\mu\nu\lambda\rho}(q) = -i
  \int\!d^4x\,e^{-iq\cdot x}\,
  \theta(x^0)[T^{\mu\nu}(x),\,T^{\lambda\rho}(0)] \,. 
\end{equation}
To see that, let us first show that $G_R^{t\nu\lambda\rho}(\omega,{\bf
0})=0$.  Indeed, according to the definition~(\ref{GRdef}),
\begin{equation}\label{GRq=0}
  G_R^{t\nu\lambda\rho}(\omega,{\bf 0})=-i\int\!dt\,e^{i\omega t}
  \<[P^\nu,\, T^{\lambda\rho}(0)]\> \,,
\end{equation}
where $P^\nu=\int\!d^3\x\,T^{t\nu}(t,\x)$ is the conserved momentum.
Now since $-i[P^\nu,\,T^{\lambda\rho}]=\d^\nu T^{\lambda\rho}$, the
right hand side of eq.~(\ref{GRq=0}) vanishes if translational
invariance is respected (as it is for a plasma in thermal equilibrium).
Now, by putting $q=0$, $\omega\neq0$ into eqs.~(\ref{Ward1}), one
immediately sees that $G$ cannot coincide with the retarded Green
function $G_R$.  Rather, the two should differ by contact terms.  If
one is interested only in the regime of small $\omega$ and $q$, the
leading infrared behavior of some of these contact terms can be found,
\begin{equation}
  G^{tttt} = G^{tttt}_R - \epsilon\,,\qquad
  G^{ttij} = G^{ttij}_R + P\delta^{ij}\,,\qquad
  G^{titj} = G^{titj}_R +\epsilon\delta^{ij}\,.
\end{equation}

The restriction coming from the Ward identities is so severe that the
Green's function can be found explicitly, if one assumes, based on
hydrodynamic considerations, that the only singularities of the
Green's functions are simple poles at $\omega=\pm q/\sqrt3$. 
(For eqs.(2.15) and (2.16) below we neglect the imaginary part in the
pole (2.1), which is small compared to the real part when q is small.)  
Then the
Green's functions are
\begin{equation}\label{hydro_corr1}
  G^{\mu\nu\lambda\rho}(\omega,q) = \frac P{3\omega^2-q^2}
  P^{\mu\nu\lambda\rho}(\omega,q)\,,
\end{equation}
where $P^{\mu\nu\lambda\rho}(\omega,q)$ are polynomials of $\omega$
and $q$ that can be found by successively applying the Ward
identities,
\begin{equation}\label{hydro_corr2}
\begin{split}
  P^{tttt} &= 3(5q^2-3\omega^2)\,, \qquad
  P^{tttz} = 12\omega q\,,\qquad
  P^{ttzz} = 3(q^2+\omega^2)\,,\\
  P^{tztz} &= q^2+9\omega^2\,, \qquad ~~~~~
  P^{tzzz} = 4\omega q\,,\qquad ~
  P^{zzzz} = -q^2+7\omega^2\,,\\
  P^{ttaa} &= 6(q^2+\omega^2)\,,\qquad ~~\,
  P^{tzaa} = 8\omega q\,,\qquad ~
  P^{zzaa} = 2(q^2+\omega^2)\,,\\
  P^{aaaa} &= 16\omega^2\,.
\end{split}
\end{equation}
These expressions are valid up to corrections of higher orders in
$\omega$ and $q$ (negligible in the infrared).  We now
show that eqs.~(\ref{hydro_corr1}) and (\ref{hydro_corr2}) are
reproduced from gravity.

\section{Gravity picture}
\label{sec:gravity}

According to the gauge theory/gravity correspondence, the
near-horizon limit of the non-extremal gravitational background of $N$
black three-branes is dual to the ${\cal N}=4$ $SU(N)$ SYM at 
finite temperature in the limit $N\rightarrow \infty$,
 $g^2_{YM}N \rightarrow \infty$. 
The relevant $5d$ part of the background metric is given by
\begin{equation}
ds^2 = 
  \frac{(\pi T R)^2}u
\left( -f(u) dt^2 + dx^2 + dy^2 +dz^2 \right) 
 +{R^2\over 4 u^2 f(u)} du^2\,,
\label{near_horizon}
\end{equation}
where $f(u)=1-u^2$.  We use the same notations and conventions as in
\cite{Policastro:2002se}.  As in \cite{Policastro:2002se}, we shall
consider a small perturbation of the background (\ref{near_horizon}),
$g_{\mu\nu} = g_{\mu\nu}^{(0)}+ h_{\mu\nu}$, where we assume
$h_{\mu\nu}$ to be dependent only on $t$ and $z$.  There are three
types of perturbations (classified by the spin under the O(2)
rotations in the $xy$ plane) listed explicitly in
ref.~\cite{Policastro:2002se}.  The tensor and the vector
perturbations were treated in ref.~\cite{Policastro:2002se}, where it
was shown that the calculations for the vector perturbations concides
with the hydrodynamic expectation for the diffusive shear modes.
Here, we are interested in the scalar perturbations where the only
non-zero elements of $h_{\mu\nu}$ are $h_{tt}$, $h_{xx}=h_{yy}$,
$h_{zz}$ and $h_{tz}$.  These perturbations correspond to sound waves
in field theory.

\subsection{Linearized equations of motion}

It will be convenient to use the Fourier decomposition
\begin{equation}
h_{\mu\nu}(t,z) = \int\! {d\omega\, d q\over (2\pi )^2} 
e^{-i\omega t + i q z}
h_{\mu\nu} (\omega,q,u)\,,
\label{fourier}
\end{equation}
and to introduce the dimensionless energy and momentum,
\begin{equation}
  \wn = \frac\omega{2\pi T}\,, \qquad \qn = \frac q{2\pi T} \,.
\end{equation}

We work in the gauge $h_{u\mu}=0$ for all $\mu$ (including $\mu=u$).
Then, to the first order in perturbation, the Einstein equations read
\begin{subequations}\label{grav}
\begin{equation}
H_{tt}'' -{3 u\over f} H_{tt}' - H_{ii}''+ {u\over f}H_{ii}'  = 0 \,,
\end{equation}
\begin{equation}
 \wn \Bigl( H_{ii}' + {u\over f} H_{ii}\Bigr) + \qn 
 \Bigl( H_{tz}' + {2 u\over f} H_{tz}\Bigr) = 0\,,
\end{equation}  
\begin{equation}
 \qn  \left( fH_{tt}' - u H_{tt}\right) + \wn H_{tz}' 
- \qn f H_{aa}'  = 0\,,
 \end{equation}
 \begin{equation}
 H_{tt}'' - {1+u^2\over 2 u f}(3 H_{tt}' - H_{ii}') 
- {1 \over u f^2}(\qn^2 f H_{tt} + \wn^2 H_{ii} 
+ 2 \wn\qn H_{tz}) =0\,,
 \end{equation}
 \begin{equation}
 H_{tz}'' - {H_{tz}'\over u} + {\wn\qn\over uf} H_{aa} = 0\,,
  \end{equation}
 \begin{equation}
 H_{aa}'' - {2\over u f} H_{aa}' + {H_{tt}'-H_{zz}'\over u}
+  {\wn^2 - \qn^2 f\over u f^2} H_{aa}  =0\,,
 \end{equation}
 \begin{equation}
 H_{zz}'' - {3{+}u^2\over 2 u f}H_{zz}' + {H_{tt}'{-}H_{aa}'\over 2 u}
 + {\wn^2 H_{zz}+ 2\wn\qn H_{tz} + \qn^2f(H_{tt}{-}H_{aa})\over u f^2}
= 0, 
 \end{equation}
\end{subequations}
where we have defined $H_{tt}=u h_{tt}/f (\pi T R)^2$, 
$H_{tz}=u h_{tz}/(\pi T R )^2$, 
$H_{ij}= u h_{ij}/(\pi T R)^2$, 
$H_{aa}=H_{xx}+ H_{yy}$, $H_{ii}= H_{aa}+H_{zz}$.
 
Not all equations in the system (\ref{grav}) are independent.  This
can be seen already from the fact that the number of equations (7)
is larger than the number of unknown functions (4).  Indeed, all
equations in (\ref{grav}) can be derived from the following system of
four equations
\begin{subequations}\label{igrav}
\begin{eqnarray}
&&H_{tt}'' - \frac{3u}{f} H_{tt}' -  H_{ii}''   + \frac{u}{f}
H_{ii}'  = 0     \,,
 \stru \\
&&\wn \Bigl( H_{ii}' +  \frac{u}{f} H_{ii}\Bigr) + \qn \Bigl( 
H_{tz}' + \frac{2 u}{f}
H_{tz}\Bigr) = 0 \,, \struu \\
&&  \qn  \left( 
 fH_{tt}' - u H_{tt}\right) + \wn H_{tz}' - \qn f H_{aa}' =
0 \,, \stru \\
&&H_{ii}'  -  \frac{ 3\,  f H_{tt}'}{3-u^2} 
 - \frac 2{f(3-u^2)}\left[
  \wn^2  H_{ii} + 2 \wn\qn H_{tz} + \qn^2f(H_{tt}-H_{aa})\right]
  = 0 \,.
\end{eqnarray}
\end{subequations}
We have checked that the left hand sides of eqs.~(\ref{grav}) are
linear combinations of the left hand sides of eqs.~(\ref{grav}) and
their derivatives.  The number of integration constants for
eqs.~(\ref{igrav}) is 5, which corresponds to four Dirichlet boundary
conditions at $u=0$ and one condition of incoming waves at $u=1$.

Equations~(\ref{grav}) (and hence (\ref{igrav})) are invariant under
residual gauge transformations, i.e., those gauge transformations
which do not break the gauge choice $h_{u\mu}=0$.  These residual
gauge transformations are written explicitly in the Appendix.  By
performing the residual gauge transformations on the trivial solution
$H_{\mu\nu}=0$, one can obtain the pure-gauge solutions to
eqs.~(\ref{igrav}).  These pure-gauge solutions are linear combinations
of $H^I$, $H^{II}$ and $H^{III}$ whose explicit forms are (only nonzero
components are written down)
\begin{subequations}\label{pure_I}
\begin{eqnarray}
H_{tz}^I &=& \wn\,,   \\
H_{zz}^I &=& -2\, \qn\,,  
\end{eqnarray}
\end{subequations}
\begin{subequations}\label{pure_II}
\begin{eqnarray}
H_{tt}^{II} &=& -2\, \wn\,,  \\
H_{tz}^{II} &=& \qn\, f\,,
\end{eqnarray}
\end{subequations}
\begin{subequations}\label{pure_III}
\begin{eqnarray}
H_{tt}^{III} & =& { 1+u^2+2\wn^2 u\over \sqrt{f}}
\,,  \stru  \\
H_{tz}^{III} & =& - \qn \, \wn \, \arcsin{u}
  -   \qn \, \wn \, u \, \sqrt{f}\,,  \stru  \\
H_{aa}^{III} & =&  - 2 \,  \sqrt{f}\,,
  \stru  \\ 
H_{zz}^{III} & =&  2\, \qn^2 \, \arcsin{u} -  \sqrt{f}  \,.
\end{eqnarray}
\end{subequations}
As we shall see below, the two remaining independent solutions
correspond to the incoming and outgoing waves.


Our first step towards solving eqs.~(\ref{igrav}) is to determine the
behavior of the solution near the horizon $u=1$.  For this end, it is
useful to temporarily abandon eqs.~(\ref{igrav}) and start again from
eqs.~(\ref{grav}), which we rewrite as a system of six first-order
differential equations,\footnote{The sixth spurious equation is needed
to avoid an irregular singularity.  The spurious solution can be
eliminated at the end by a direct check.}
\begin{subequations}\label{iigrav}
\begin{eqnarray}
  &&H_{tt}' = {1\over f} P_{tt}\,, \stru  \\
  &&H_{aa}' = {1\over f} P_{tt}- {u\over f} H_{tt} 
             + {\wn \over \qn f} P_{tz}\,,\stru   \\
  &&H_{ii}' = - {u\over f} H_{ii} - {\qn \over \wn}  P_{tz} -
 {2 \qn u\over \wn f}  H_{tz}\,, \struu  \\
  &&H_{tz}' = P_{tz} \,, \stru  \\
  &&P_{tz}' = {1\over u} P_{tz} - {\wn \qn \over u f} H_{aa}
 \,,  \stru  \\
  &&P_{tt}' = - {u^2{-}3\over 2 u f} P_{tt} + {\qn^2\over u} H_{tt} 
 + {\qn (1{+}u^2) \over 2 \wn u} P_{tz}
 +
 {u{+}u^3{+}2\wn^2\over
 u f} \left[\frac{H_{ii}}2 + \frac\wn\qn H_{tz}\right]\,. 
\end{eqnarray}
\end{subequations}
In matrix notation, eqs.~(\ref{iigrav}) read
\begin{equation}
X' = A(u)\,  X\,,
\label{matrix}
\end{equation}
where $X^T=(H_{tt},H_{aa},H_{ii},H_{tz},P_{tz},P_{tt})$, and $A[u]$ is
a $6\times 6$ matrix which is singular at the horizon $u=1$.

To find the indices 
characterizing the behavior of the solution there, one substitutes the ansatz
$X = (1-u)^r F[u]$ into (\ref{matrix}) and matches the coefficients 
of the leading singular terms on both sides. This amounts to finding 
the eigenvalues of the matrix $\bar{A}= - \lim_{u\rightarrow 1} (1-u) A[u]$.
The six eigenvalues are $r_{1}= r_2 = 0$, $r_{3} = - 1/2$, 
$r_{4} =  i\wn /2$,
 $r_{5} = - i\wn /2$, $r_{6} = 1/2$ . 
Five of the corresponding eigenvectors are given by
$F_1=(0,0,-2\qn,\wn,0,0)$, $F_2=(\wn,0,0,0,\qn,0)$, $F_3=(1,0,0,0,0,1)$,
$F_4=(0,i,0,0,\qn,0)$, $F_5=(0,-i,0,0,\qn,0)$. Then a linear 
combination of $X_k = (1-u)^{r_k} F_k$, $k=1,...5$,
represents a local solution near the horizon.
The sixth eigenvector, $F_6=(1+\wn^2,2,2,0,2\wn\qn,-1-\wn^2)$,
 is a spurious one
(one can check that $\sqrt{1-u} F_6$
  is not a local solution of eqs.~(\ref{igrav})), 
and should be discarded.
The eigenvectors with $r=0,0,-1/2$ correspond to
 the three independent pure gauge 
solutions given explicitly in eqs.~(\ref{pure_I}),
(\ref{pure_II}) and
(\ref{pure_III}), respectively.  The
solutions with indices $\mp i\wn /2$ represent incoming/outgoing waves.

Having found the indices and local solutions near the horizon, we can
now return to eqs.~(\ref{igrav})
and solve them perturbatively in the hydrodynamic regime $\wn \ll 1$ 
and $\qn \ll 1$. 
Our knowledge of the local behavior of the solutions near 
the horizon allows us to isolate the solution corresponding to 
the incoming wave. To the second order in $\wn$ and $\qn$ it is given by 
$H^{inc}=(H^{inc}_{tt}, H^{inc}_{ii}, H^{inc}_{tz}, H^{inc}_{aa})$, where 
\begin{subequations}
\begin{eqnarray}
&&H^{inc}_{tt}  = \, \frac{\qn^2}{3} \left( 1-u \right) \, , \struu \\
&&H^{inc}_{ii}  = \,\qn^2 (1-u) \, , \stru \\
&&H^{inc}_{tz}  =  \, - \frac{i\qn}{2}\, \left( 1-u^2 \right) 
+ \frac{\wn\qn}{2}\, u \left( 1-u \right) -
\frac{\wn\qn}{4} \left( 1-u^2 \right) \ln \frac{2\,(1-u)}{1+u} \, ,\stru\\ 
&&\begin{split}
  H^{inc}_{aa}  &= \, 1 - \frac{i\wn}{2} 
\ln\frac{1{-}u^2}{2} -
\frac{\wn^2}{8} \ln^2 (1{-}u) - \frac{\wn^2}{4} \ln (1{-}u)
\ln \frac{1{+}u}2 \stru \\
&\quad+ \frac{2 \qn^2}{3} ( 1{-}u )+ {3\wn^2 {+} \qn^2 \over 3} \ln
  \frac{1{+}u}2 + \frac{\wn^2}{8} \ln^2
\frac{1{+}u}2
 - \frac{\wn^2}{2} \, \textrm{Li}_2  \frac{1{-}u}{2} \,.
\end{split}
\end{eqnarray}
\end{subequations}
Thus, to the second order in $\wn$ and $\qn$, the most general
solution to eqs.~(\ref{igrav}) which satisfies the incoming-wave
boundary condition at the horizon is given by
\begin{equation}
H(u) = a\, H^{inc}(u) + b\, H^I (u) +  c\, H^{II} (u) + 
d\, H^{III} (u)\,,
\label{solution}
\end{equation}
where $a$, $b$, $c$ and $d$ are constants, and $H^{I,II,III}$ are the
pure-gauge solutions given in eqs.~(\ref{pure_I}), (\ref{pure_II}) and
(\ref{pure_III}).  The constants $a$, $b$, $c$ and $d$ are determined
from the Dirichlet conditions at the boundary $u=\epsilon \rightarrow
0$ and can be expressed in terms of the boundary values $H_{tt}^{0}$,
$H_{aa}^{0}$, $H_{ii}^{0}$, $H_{tz}^{0}$.

\subsection{The action}
\label{sec:action}

 The action is given by the sum of three terms,
\begin{equation}
S = S_M + S_{\partial M}^{(1)} +  S_{\partial M}^{(2)}\,,
\end{equation}
where $S_{\partial M}^{(1)}$ is the Gibbons-Hawking boundary term, 
and  $S_{\partial M}^{(2)}$ is proportional to the volume of the boundary.
Explicitly,
\begin{equation}
S = 
{\pi^3 R^5\over 2 \kappa^2_{10}}\left[ \;  
  \int\limits_{0}^{1}\!du\, d^4x\, \sqrt{-g}
\left( {\cal R} - 2\Lambda\right) + 2 \int\! d^4x\, \sqrt{-h} K + 
a \int\! d^4x\, \sqrt{-h} \; \right]\,.
\label{action_grav}
\end{equation}
Here $\kappa_{10} = \sqrt{8\pi G}$ is the ten-dimensional
 gravitational constant,
related to the parameter $R$ of the non-extremal geometry and the number
$N$ of coincident branes by 
$\kappa_{10} = 2\pi^2\sqrt{\pi}R^4/ N$~\cite{Gubser:1996de}. 
Following ref.~\cite{Liu:1998bu}, we choose $a=-6/R$ to cancel
the volume divergence in eq.~(\ref{action_grav}).
On shell, the action reduces to the surface terms, $S=
S_{horizon} + S_{\epsilon}$, where
\begin{equation}\label{action_onshell}
\begin{split}
S_{\epsilon} &= {\pi^2 N^2 T^4\over 8} \int\! d^4x\,  \Biggl[ - 1  +
 {1\over 2}\left( 3 H_{tt} + H_{ii}\right) \\
 &\quad+
{1\over 8}\left( 3 H_{tt}^2  - 12 H_{tz}^2 + 2 H_{tt}H_{ii}
  +  2 H_{zz}H_{aa} - H_{zz}^2 \right)\\
 &\quad- {1\over 2 \epsilon}\left( 
  H_{tz}^2 + \frac14 H_{aa}^2
  -  H_{tt}H_{ii}
   + H_{zz}H_{aa}\right)' \Biggr]\,.
\end{split}
\end{equation}
We can now substitute our solution (\ref{solution}) into 
eq.~(\ref{action_onshell}) and compute the correlators. At leading order, 
we retain only terms linear in $\wn$ and $\qn$ in eq.~(\ref{solution}). 
We have\footnote{Terms quadratic in $H$ in 
(\ref{action_onshell_0}) should be understood as products 
$H(\omega,\q)H(-\omega,-\q)$,
and integration over $\omega$ and $\q$ is implied.}
\begin{eqnarray}
S_{\epsilon} &=& {\pi^2 N^2 T^4\over 8 } \Biggl[  - 
 V_4  +
 {1\over 2}\left( 3 H_{tt}^0(0) +  H_{ii}^0(0)\right) \nonumber  \\
&&+ {1\over 2 (\qn^2 {-} 3 \wn^2)}\left(
\left( 2 \qn H_{tz}^0 {+} \wn H_{ii}^0 \right)^2 + H_{tt}^0 
\Bigl(3\qn^2 H_{tt}^0 {+}
12 \qn\wn H_{tz}^0  {+} (3 \wn^2{+}\qn^2) H_{ii}^0
 \Bigr)\right) \nonumber\\
&&+ \frac18 \Bigl(3 (H_{tt}^0)^2 - 
 12 (H_{tz}^0)^2 +  2 H_{tt}^0 H_{ii}^0+ 
  2 H_{zz}^0 H_{aa}^0 - 
 (H_{zz}^0)^2\Bigr)\Biggr]\,.
\label{action_onshell_0}
\end{eqnarray}
The constant term, $-\pi^2 N^2 T^4V_4/8$, is the free energy density
(i.e., the pressure with a minus sign)~\cite{Gubser:1998nz}, times the
four-volume.

\subsection{The correlators}

As noticed in ref.~\cite{gamma_paper}, in general it is not possible
to calculate the Minkowski two-point Green's function by
differentiating the gravitational action with respect to the boundary
values of fields, since the result of this differentiation is necessarily
real while the retarded Green's function is in general complex.  
In other words, in Minkowski AdS/CFT, the boundary action in general fails to 
be a generating functional for the Green's functions, and a prescription 
formulated in  ref.~\cite{gamma_paper} should be used to obtain 
the correct results. 
We remark, however, that to leading order this problem does not
arise, because the correlators are real (cf. eqs.~(2.15) and (2.16)).
Thus $S_\epsilon$ in eq.~(3.16) can be used as the generating functional
to this order.%
\footnote{As discussed in  ref.~\cite{gamma_paper}, the surface 
term $S_{horizon}$ must be dropped.}
The coefficient of
proportionality is easily determined by using the naive equality
\begin{equation}
\langle e^{i \int \phi_0 {\cal O}} \rangle = e^{i S_{\epsilon}}\,.
\label{heuristic}
\end{equation}
In our case, the boundary value of the metric perturbation $h^i_j$
acts as a source for the stress-energy tensor~\cite{Liu:1998bu}, with
the coupling given by
\begin{equation}
 \frac{1}{2}\int dt \, d^3 x h_\mu^\nu T_\nu^\mu  =  
\frac{1}{2} \int dt \, d^3 x \left( H_{tt}^0 T^{tt} + \frac{1}{2}
H_{aa}^0 T^{aa} + H_{zz}^0 T^{zz} + 2 H_{tz}^0 T^{tz} \right) \,.
\label{coupling}
\end{equation}

Taking the derivatives of $S_{\epsilon}$ with the normalization
ensured by (\ref{coupling}), one finds the following one-point
functions
\begin{subequations}
\begin{eqnarray}
\left\<  T^{tt}  \right\> &=& 2 {\delta  S_{\epsilon}\over \delta H_{tt}^0}
 = {3  \pi^2 \over 8} N^2 T^4\,,\label{energy} \stru \\
\left\<  T^{aa}  \right\> &=& 4 {\delta  S_{\epsilon}\over \delta H_{aa}^0}
 = {  \pi^2 \over 4}N^2 T^4\,,\label{pressure_aa} \stru \\
\left\<  T^{zz}  \right\> &=& 2 {\delta  S_{\epsilon}\over \delta H_{zz}^0}
 = {  \pi^2 \over 8} N^2 T^4 \,.\label{pressure_zz} 
\end{eqnarray}
\end{subequations}
These results are in complete agreement with the known thermodynamics
of our theory: indeed, $\<T^{tt}\>=\epsilon=3P$,
$\<T^{aa}\>=\<T^{xx}+T^{yy}\>=2P$ and $\<T^{zz}\>=P$ with the pressure
obtained in ref.~\cite{Gubser:1996de}, $P = \pi^2 N^2 T^4/8$.


The two-point functions easily follow from $S_{\epsilon}$ by taking
into account eq.~(\ref{heuristic}) and eq.~(\ref{coupling}).  The
results are in complete agreement with hydrodynamics,
eqs.~(\ref{hydro_corr1}) and (\ref{hydro_corr2}).  For example,
\begin{equation}
  G^{tttt}(\omega,q) = 
- 4 {\delta^2  S_{\epsilon}\over \delta (H_{tt}^0)^2} = 
{3N^2\pi^2 T^4 q^2\over 2 (3\omega^2 - q^2)} 
- {3  \pi^2 \over 8} N^2 T^4 
  = 3P\frac{5q^2-3\omega^2}{3\omega^2-q^2}\,.
\end{equation}
Other correlators are computed similarly.  The singularity
structures, and more unexpectedly, all contact terms are correctly
reproduced by gravity.  These formulas are valid up to the corrections
of order $O(\wn^2,\qn^2,\wn\qn)$ (the corrections are discussed below
in section~\ref{attenuation}).  We observe that the correlation
functions exhibit a pole corresponding to the propagation of the sound
wave in the hot ${\cal N}=4$ SYM plasma, with the correct value for
the speed of sound, $v_s = 1/\sqrt{3}$.  It is truly amazing that the
{\it quantitative} prediction of field theory follows from the 5d
non-supersymmetric gravity background. This result boosts our
confidence in the validity of the gauge/gravity correspondence in
general.

Since the Green's functions coincide with the forms expected from
hydrodynamics, they automatically satisfy the Ward
identities~(\ref{Ward1})--(\ref{Ward3}).  This is the consequence of
the invariance of the boundary gravitational action (\ref{action_onshell_0}),
which serves as the generating functional,\footnote{With the reservations 
discussed above.} under the gauge
transformations (\ref{gauge_I}), (\ref{gauge_II}) and
(\ref{gauge_III}).

\subsection{Sound attenuation}
\label{attenuation}

To find the imaginary part of the dispersion
relation~(\ref{hydro_sound}), one has to perform calculation beyond
leading order.  That is done by using the full
solution~(\ref{solution}) and keeping the terms quadratic in $\wn$ and
$\qn$.  We have to follow the prescription of  ref.~\cite{gamma_paper}, 
since to this order in $\wn$, $\qn$ the boundary action  $S_{\epsilon}$ 
 is no longer a generating functional for the Green's functions.%
\footnote{We have not checked the 
Ward identities in this approximation.}

\begin{figure}[h]
\begin{center}
\epsffile{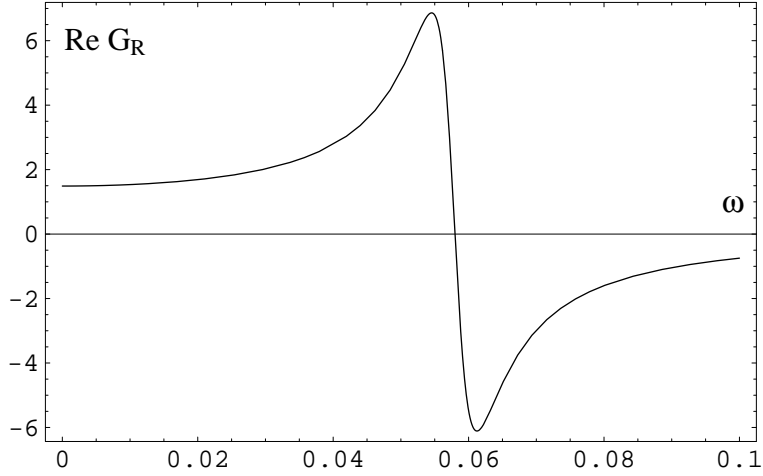}
\end{center}
\caption{The real part of the correlator $G^R_{tt,\,tt}(\wn,\qn)$ 
as a function of (real) $\wn$ at $\qn = 1/10$.  
}
\label{reG}
\end{figure}

\begin{figure}[h]
\begin{center}
\epsffile{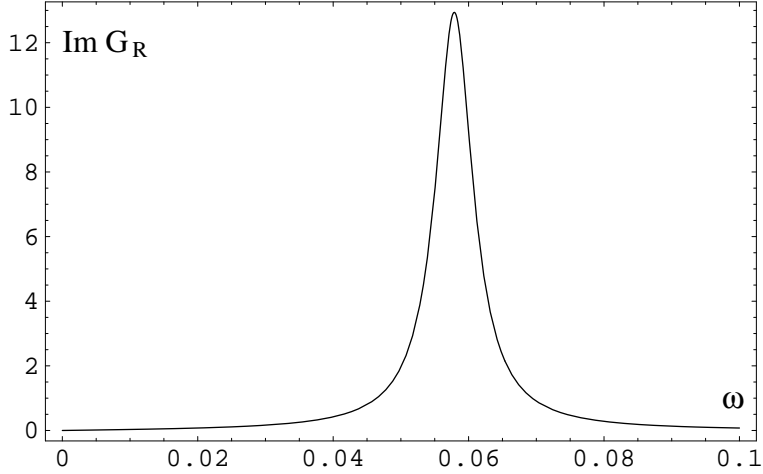}
\end{center}
\caption{The imaginary part of the correlator $G^R_{tt,\,tt}(\wn,\qn)$ 
as a function of (real) $\wn$ at $\qn = 1/10$. The peak corresponds to the 
(attenuated) pole at $\Re \wn = \qn/\sqrt{3} \approx 0.0577$.
}
\label{imG}
\end{figure}

The resulting analytic expressions for the correlators are
rather cumbersome, and will not be presented here.  The real and
imaginary parts of a typical correlator are plotted in
figs.~\ref{reG},\ref{imG}.  We will discuss only the location of the
pole to this next-to-leading order.  It is now given by the equation
\begin{equation}
\wn^2 - {\qn^2 \over 3} + {2 i\over 3}\, \wn \, \qn^2 - {i \wn \log{2}\over 6}
\left( \qn^2 - 3 \, \wn^2 \right) = 0\,.
\label{the_pole}
\end{equation}
Near the pole we have $\wn \approx \qn/\sqrt{3}$. Thus, in this region
the last term in (\ref{the_pole}) is of order $\wn^2\qn^2$, and can be
neglected.  We obtain
\begin{equation}
\wn = {\qn\over \sqrt{3}} - {i \qn^2\over 3} + O (\qn^3)\,,
\end{equation}
or, in terms of original variables,
\begin{equation}
\omega = {q\over \sqrt{3}} - {i q^2\over 6 \pi T} + O (q^3)\,.
\end{equation}
This is in complete agreement with hydrodynamics, which predicts (see
eq.~(\ref{hydro_sound}) and recall that $\zeta=0$ in conformal
theories) that the imaginary part of the sound-wave dispersion
relation is
\begin{equation}
  -\frac{2i}3  \frac\eta{\epsilon+P} q^2 = -\frac{iq^2}{6\pi T}\,,
\end{equation}
where we have used the value of $\eta/(\epsilon+P)=(4\pi T)^{-1}$
computed in refs.~\cite{viscosity,Policastro:2002se}.


\section{Conclusion}
\label{sec:concl}
In this paper we have demonstrated how sound waves emerge from
gravity, with a sound speed in agreement with field-theoretical
expectations.  Together with the result of our previous
work~\cite{Policastro:2002se} we have found that the AdS/CFT
correspondence is capable of reproducing the hydrodynamic regime of
the correlators in $\N=4$ SYM theory at finite temperature, which is
another testimony to the validity of the correspondence.

The calculations of ref.~\cite{Policastro:2002se} has been extended in
ref.~\cite{Herzog} to M2 and M5 branes of M-theory, where hydrodynamic
behavior is also observed, although the worldvolume theories in these
cases are not well understood.\footnote{Methods for computing the 
absorption cross section by non-extremal $D$- and $M$-branes 
were discussed in detail in \cite{Policastro:2001yb}.}
 It should be possible to find the
sound waves in these cases too.

It would be interesting to investigate whether the hydrodynamic
behaviors observed so far are generic for all black $p$-branes
solutions, i.e., are inherent properties of general relativity (as,
for example, the entropy), or are specific properties of metrics which
have dual field-theoretical description.

\begin{acknowledgments}
The authors thank L.G.~Yaffe for
discussions and K.~Skenderis for correspondence.
Our special thanks to Massimo Porrati for resolving an 
apparent puzzle 
we had encountered along the way. G.P. would like to thank
INT for 
its hospitality.  
  This work is supported, in part, by DOE Grant No.\
DOE-ER-41132.  The work of D.T.S.\ is supported, in part, by the Alfred P.\
Sloan Foundation. G.P. is supported by PPARC. 
\end{acknowledgments}

\appendix

\section{Gauge transformations}
\label{sec:gauge}

The requirement that the infinitesimal diffeomorphism
\begin{eqnarray}
  x^{\mu} & \rightarrow & x^{\mu} + \xi^{\mu}\,,\\
  g_{\mu\nu} & \rightarrow & g_{\mu\nu} - \nabla_{\mu}\xi_{\nu} - 
 \nabla_{\nu}\xi_{\mu}
\label{gauge_metric_transf}
\end{eqnarray}
preserves the gauge conditions $h_{u\mu}=0$ leads to the equation 
\begin{equation}
\partial_{\mu} \xi_{u} + \partial_{u} \xi_{\mu} -
 2 \Gamma_{\mu u}^{\rho}\xi_{\rho} =0\,,
\label{gauge_eq}
\end{equation} 
which constrains possible choices of $\xi_{\mu}$ for residual gauge
transformations.  Since the covariant derivative in
eq.~(\ref{gauge_metric_transf}) is taken with respect to the full
metric, the residual gauge transformations are dependent on the
fluctuations $h_{\mu\nu}$.  It will be useful to introduce a
book-keeping parameter $\kappa$ associated with the fluctuations
$h_{\mu\nu}$ and expand both $\xi_{\mu}$ and the Christoffel symbols
in series over $\kappa$,
\begin{eqnarray}
 \xi_\mu &=& \xi_\mu^{(0)} + \kappa \, \xi_{\mu}^{(1)}+\cdots\,,\\
\Gamma &=& \Gamma^0 + \kappa \, \Gamma^{(1)}+\cdots\,.
\end{eqnarray}
Then, to the linear order in fluctuations, a general residual gauge
transformation is linear combinations of the following three types of
transformations:

1) The transformations generated by
\begin{subequations}
\begin{eqnarray}
  \xi_z &=&  {C_z(t,z)\over u} +  \xi_{z}^{(1)}(u,t,z)\,, \stru \\
  \xi_t &=&  \xi_{t}^{(1)}(u,t,z)\,,
\end{eqnarray}
\end{subequations}
where $ \xi_{z}^{(1)}$ and $ \xi_{t}^{(1)}$ satisfy, respectively,
\begin{subequations}
\begin{eqnarray}
\partial_{u} \xi_{z}^{(1)} +{1\over u} \xi_{z}^{(1)}  &=&
{C_1 H_{zz}'\over u}\,,\\
\partial_{u} \xi_{t}^{(1)}  + {1+u^2\over u f} \xi_{t}^{(1)} &=&
 {f C_1\over u}
 \left( {H_{tz}\over f}\right) '\,, 
\end{eqnarray}
\end{subequations}
They are given by $H_{\mu\nu}\rightarrow H_{\mu\nu} +
H_{I\,\mu\nu}^{gauge}$, where the only nonzero terms in
$H_{I\,\mu\nu}^{gauge}$ are\footnote{In all formulas in this Appendix 
products involving $H_{ij}$ should be understood as convolutions.
For example, $(\omega\, C_1\, H_{zz}) (\omega,q) = \int  \omega'\,
C_1(\omega', q')\, H_{zz} (\omega - \omega', q-q') \, d\omega' d q'$.} 
\begin{subequations}\label{transf_I}
\begin{eqnarray}
H_{I\, tt}^{gauge} &=& {2\wn u\over f} \xi_t^{(1)}
 - \qn \, C_1 H_{tt} -
 {2\, \wn \, C_1\over  f } H_{tz} \,,
\label{transf_I_tt}
\stru  \\
H_{I\, tz}^{gauge} &=&  
 \wn C_1 + u\, \wn \, \xi_z^{(1)} - u \qn\, \xi_t^{(1)}
- \wn \, C_1 H_{zz}\,,\label{transf_I_tz} \stru  \\
H_{I\, aa}^{gauge} & = & - \qn \, C_1 H_{aa}\,, 
\label{transf_I_aa}
\stru  \\
H_{I\, zz}^{gauge} & = & -2 \qn C_1 - 2 \, u \, \qn \, \xi_z^{(1)} 
+ \qn \, C_1 H_{zz}\,.
\label{transf_I_zz}
\end{eqnarray}
\end{subequations}
Here and later we use $C_1 = i (2\pi T) C_z$, $C_2 = i (2\pi T) C_t$
and $C_3=i(2\pi T)C_u$.

2) The transformations generated by 
\begin{subequations}
\begin{eqnarray}
\xi_t &=& - {f C_t(t,z)\over u} +  \xi_{t}^{(1)}(u,t,z)\,, \stru \\
\xi_z &=&  \xi_{z}^{(1)}(u,t,z)\,,
\end{eqnarray}
\end{subequations}
where $ \xi_{t}^{(1)}$ and $ \xi_{z}^{(1)}$ obey the following
inhomogeneous equations
\begin{subequations}
\begin{eqnarray}
\partial_{u} \xi_{t}^{(1)}  + {1+u^2\over u f} \xi_{t}^{(1)}  &=&
 {f C_2\over u }  H_{tt}' \,,\\
\partial_{u} \xi_{z}^{(1)} +{1\over u} \xi_{z}^{(1)} &=&
{C_2 H_{tz}'\over u}\,.
\end{eqnarray}
\end{subequations}
To the linear order in $\kappa$, the transformations are 
given by  $H_{ij}\rightarrow H_{ij} + H_{II\, ij}^{gauge}$, where
\begin{subequations}\label{transf_II}
\begin{eqnarray}
  H_{II\, tt}^{gauge}&=& -2\,\wn \, C_2 + {2\wn u\over f} \xi_t^{(1)} -
  \wn \, C_2 \, H_{tt} \,, \stru  \\
H_{II\, tz}^{gauge}& =& \qn \, f\, C_2  - u\, \qn \, \xi_t^{(1)}
 + u\,\wn \, \xi_{z}^{(1)}
+ \qn \, f \, C_2 \, H_{tt}\,, \stru  \\
H_{II\, aa}^{gauge}& =&  \wn \, C_2 \, H_{aa}\,, \stru  \\
H_{II\, zz}^{gauge}& =& -2\, u\, \qn\,  \xi_z^{(1)}
+ 2\, \qn \, C_2\, H_{tz} + \wn \, C_2 \, H_{zz}\,. 
\end{eqnarray}
\end{subequations}

3) A third group of transformations is generated by
\begin{subequations}
\begin{eqnarray}
\xi_u &=& {C_u(t,z)\over u\sqrt{f}}\,, \stru \\
\xi_z &=& - \partial_z C_u(t,z) {\arcsin{u}\over u}\,,  \stru \\
\xi_t &=& - \partial_t C_u(t,z) \sqrt{f}\,,
\end{eqnarray}
\end{subequations}
and is given by $H_{ij}\rightarrow H_{ij} + H_{III\, ij}^{gauge}$,
where
\begin{subequations}\label{transf_III}
\begin{eqnarray}
&&\begin{split}
 H_{III\, tt}^{gauge} &= {C_3\, (1+u^2+2\wn^2 u)\over \sqrt{f}} +
 {C_3\, u^2\over \sqrt{f}} \left( {f H_{tt}\over u}\right)' \\
 &\quad+
 {C_3\, u \, \wn^2 \, H_{tt}\over \sqrt{f}} +
 {C_3\, \arcsin{u} \over  f} \left( \qn^2 f H_{tt} + 
2 \, \qn \, \wn \, H_{tz}\right)\,, 
\end{split}\stru  \\
&&\begin{split}
H_{III\, tz}^{gauge}& = - C_3\, \qn \, \wn \, 
\left( 1 -H_{zz}\right) \arcsin{u}
 + C_3 \, u^2 \, \sqrt{f} \, \left( {H_{tz}\over u}\right)'\\ 
&\quad-  C_3\,  \qn \, \wn \, u \, 
\sqrt{f} \, \left( 1 + H_{tt}\right)\,,  
\end{split}\stru  \\
&&\begin{split}
H_{III\, aa}^{gauge}& =  - 2 \, C_3 \, \sqrt{f}- 
{ C_3 \, u \, \wn^2 \, H_{aa}\over \sqrt{f}} 
+  u^2\, \sqrt{f}\,  C_3 \left( {H_{aa}\over u}\right)'  \\
&\quad+ \qn^2\,  C_3\,  H_{aa}\arcsin{u}  \,,
\end{split}\stru  \\
&&\begin{split} 
H_{III\, zz}^{gauge}& =  C_3 \, \qn^2 \, \arcsin{u}  
\left( 2 - H_{zz}\right) -C_3 \sqrt{f}+ u^2\,  
\sqrt{f} C_3 \left( {H_{zz}\over u}\right)'  \\
&\quad-  {C_3 u \over \sqrt{f}} \left( 2 \, \qn \, \wn \, 
H_{tz} +\wn^2 \, H_{zz}\right)  \,.
\end{split}
\end{eqnarray}
\end{subequations}
At the boundary, the gauge transformations to the first order in 
$\wn$, $\qn$, and to the linear order in fluctuations, reduce to 
the following three independent sets:

\noindent
Set I:
\begin{subequations}\label{gauge_I}
\begin{eqnarray}
H_{tt}^0  
&\rightarrow & H_{tt}^0 - \qn \, C_1\,  H_{tt}^0 - 2\, \wn \, C_1\, 
H_{tz}^0\,,\label{gauge_I_tt_0}  \\ 
 H_{tz}^0 &\rightarrow & H_{tz}^0  + \wn\, C_1 \, - \wn \, C_1 \, H_{zz}^0 \,,
\label{gauge_I_tz_0}  \\
 H_{aa}^0 &\rightarrow& H_{aa}^0 - \qn \, C_1 \, H_{aa}^0\,,
\label{gauge_I_aa_0}  \\
 H_{zz}^0 &\rightarrow& H_{zz}^0 - 2\, \qn \, C_1 + \qn \, C_1 \, H_{zz}^0\,.
\label{gauge_I_zz_0}
\end{eqnarray}
\end{subequations}
Set II:
\begin{subequations}\label{gauge_II}
\begin{eqnarray}
 H_{tt}^0 &\rightarrow & H_{tt}^0  -2 \wn \, C_2 -
\wn \, C_2 \, H_{tt}^0\,,\label{gauge_II_tt_0}  \\
 H_{tz}^0 &\rightarrow & H_{tz}^0 + \qn \, C_2 + \qn \, C_2 \, H_{tt}^0\,,
\label{gauge_II_tz_0}  \\
 H_{aa}^0 &\rightarrow & H_{aa}^0 + \wn \, C_2 \, H_{aa}^0 \,,
\label{gauge_II_aa_0} \\ 
 H_{zz}^0 &\rightarrow & H_{zz}^0 + 2 \qn \, C_2 \, H_{tz}^0 +
\wn \, C_2 \, H_{zz}^0\,.
\label{gauge_II_zz_0}
\end{eqnarray}
\end{subequations}
Set III:
\begin{subequations}\label{gauge_III}
\begin{eqnarray}
 H_{tt}^0 &\rightarrow & H_{tt}^0 + C_3 - C_3 H_{tt}^0 \,,
  \label{gauge_III_tt_0} \\
 H_{tz}^0 &\rightarrow& H_{tz}^0 - 
C_3 H_{tz}^0 \,, \label{gauge_III_tz_0} \\
 H_{aa}^0 &\rightarrow& H_{aa}^0 - 2 C_3 - C_3 H_{aa}^0\,,
 \label{gauge_III_aa_0}\\
 H_{zz}^0 &\rightarrow& H_{zz}^0 - C_3 - C_3 H_{zz}^0 \,.
\label{gauge_III_zz_0}
\end{eqnarray}
\end{subequations}
From the point of view of four dimensions, eqs.~(\ref{gauge_I}) and
(\ref{gauge_II}) are diffeomorphisms,  while eqs.~(\ref{gauge_III})
correspond to dilatation when $C_3$ is constant.  One can check that to
the linear order in fluctuations, the generating functional
(\ref{action_onshell_0}) is invariant under the gauge transformations
of Sets I, II and III.


\begin{thebibliography}{99}

\bibitem{Maldacena}
J.~Maldacena,
{\em The large N limit of superconformal field theories and supergravity,}
\atmp{2}{1998}{231}
[\hepth{9711200}].

\bibitem{GKP}
S.S.~Gubser, I.R.~Klebanov and A.M.~Polyakov,
{\em Gauge theory correlators from non-critical string theory,}
\plb{428}{1998}{105}
[\hepth{9802109}].

\bibitem{Witten1}
E.~Witten,
{\em Anti de Sitter space and holography,}
\atmp{2}{1998}{253}
[\hepth{9802150}].

\bibitem{Landafshitz6}
L.D.~Landau and E.M.~Lifshitz, {\em Fluid mechanics}, Pergamon Press,
New York 1987, 2nd ed.

\bibitem{gamma_paper}
D.T.~Son and A.O.~Starinets, 
{\em Minkowski space correlators in AdS/CFT correspondence: recipe 
and applications,}
\jhep{09}{2002}{42}
[\hepth{0205051}].

\bibitem{Policastro:2002se}
G.~Policastro, D.T.~Son and A.O.~Starinets,
{\em From AdS/CFT correspondence to hydrodynamics,}
\jhep{09}{2002}{43}
[\hepth{0205052}].

\bibitem{Yaffe} L.G.~Yaffe, {\em Hydrodynamic fluctuations in 
relativistic theories} (1992), unpublished.

\bibitem{Gubser:1996de}
S.S.~Gubser, I.R.~Klebanov and A.W.~Peet,
{\em Entropy and temperature of black 3-branes,}
\prd{54}{1996}{3915}
[\hepth{9602135}].

\bibitem{Liu:1998bu}
H.~Liu and A.A.~Tseytlin,
{\em $D = 4$ super Yang-Mills, $D = 5$ gauged supergravity and 
$D = 4$ conformal  supergravity,}
\npb{533}{1998}{88}
[\hepth{9804083}].

\bibitem{Gubser:1998nz}
S.S.~Gubser, I.R.~Klebanov and A.A.~Tseytlin,
{\em Coupling constant dependence in the thermodynamics of $\N = 4$  
supersymmetric Yang-Mills theory,}
\npb{534}{1998}{202}
[\hepth{9805156}].

\bibitem{viscosity}
G.~Policastro, D.T.~Son and A.O.~Starinets, 
{\em Shear viscosity of strongly coupled ${\cal N} = 4$ supersymmetric 
Yang-Mills plasma,}
\prl{87}{2001}{081601}
[\hepth{0104066}].

\bibitem{Herzog}
C.~P.~Herzog,
``The hydrodynamics of M-theory,''
JHEP {\bf 0212}, 026 (2002)
[arXiv:hep-th/0210126].

\bibitem{Policastro:2001yb}
G.~Policastro and A.~Starinets,
{\em On the absorption by near-extremal black branes,}
\npb{610}{2001}{117}
[\hepth{0104065}].



\end{thebibliography}
\end{document}